\documentclass[aps,prb,twocolumn,showpacs,superscriptaddress,longbibliography]{revtex4-2}

\usepackage{graphicx}  
\usepackage{subfigure}
\usepackage{multirow}
\usepackage{parskip}
\usepackage[T1]{fontenc}
\usepackage{amsfonts}  
\usepackage{amsmath}   
\usepackage{amssymb}   

\usepackage{bm}
\usepackage{textcomp}
\usepackage{color}
\usepackage{longtable}
\usepackage{dcolumn}
\usepackage[a4paper=true,pagebackref=false]{hyperref}

\usepackage[normalem]{ulem}
\usepackage{hyperref}
\usepackage[english]{babel}

\begin{document}
\title{Spin orbital reorientation transitions induced by magnetic field}

\author{Dariusz~Sztenkiel} \email{sztenkiel@ifpan.edu.pl}
\affiliation{Institute of Physics, Polish Academy of Sciences, Aleja Lotnikow 32/46, PL 02-668 Warszawa, Poland}

\date{\today}

\begin{abstract}  

Here we report on a new effect similar to the spin reorientation transition (SRT) that takes place at two magnetic fields of $B_{SORT1}$ and $B_{SORT2}$. The effect is observed in the magnetization curves of small Mn$^{3+}$ magnetic clusters in the wurtzite GaN (being in a paramagnetic state) calculated using crystal field model approach. The obtained results suggest that the computed magnetic anisotropy (MA) reverses its sign on increasing $B$ across $B_{SORT}$. Detailed analysis show however that MA is unchanged for high magnetic fields. We show that the observed effect arises from the interplay of the crystalline environment and the spin–orbit coupling $\lambda \hat{\textbf{S}} \hat{\textbf{L}}$, therefore we name it spin orbital reorientation transition (SORT). The value of $B_{SORT1}$ depends on the crystal field model parameters and the number of ions $N$ in a given cluster, whereas $B_{SORT2}$ is controlled mostly by the magnitude of the spin-orbit coupling $\lambda$. The explanation of SORT is given in terms of the spin $M_S$ and orbital momentum $M_L$ contributions to the total magnetization $M = M_S + M_L$. The similar effect should also be present in other materials with not completely quenched (non zero) orbital angular momentum $L$, an uniaxial magnetic anisotropy and the positive value of $\lambda$.

%


\end{abstract}

\maketitle


\section{Introduction}

The main sources that influence the magnetic anisotropy (MA) of materials are long range dipole-dipole interaction (shape anisotropy), short range exchange interaction, surface and magnetocrystalline anisotropy. The relative strength of each term can be to some degree controlled by the temperature $T$, applied magnetic field $B$, external pressure, magnetic layer thicknesses or electric field $E$, leading to the modification of MA. In particular, a spin reorientation transition (SRT) \cite{Horner:1968_PRL, Belov:1976_SovPhysUsp} is the effect in which a preferred (easy) axis of magnetization changes its orientation as a result of varying one of these driving parameters, for example $T$ \cite{Sawicki:2003_JSNM, Dmitriev:2015_LTP, Zhao:2015_PRB, Pati:2017_PSS_RRL, Hou:2020_JPCC}, $B$ \cite{Pati:2017_PSS_RRL, Oinuma:2020_PRB, Dmitriev:2015_LTP, Durbin:1975_JPhysC_SSP}, $E$ \cite{Chiba:2008_N}, sample thicknesss \cite{Miao:2013_SciChinaPMA} or carrier concentration \cite{Sawicki:2005_PRB}. Here we report on the effect similar to SRT in the dilute magnetic semiconductor (Ga,Mn)N at the magnetic field of $B_{SORT1} \cong 10$~T and $B_{SORT2} \cong 220$~T. We numerically study the paramagnetic properties of small Mn$^{3+}$ magnetic clusters in the wurtzite GaN using crystal field model (CFM) approach. The model includes the crystal field splittings, Zeeman effect, spin-orbit coupling and ferromagnetic superexchange interaction between Mn ions. The magnetization $M$ is computed as a function of the magnetic field $\textbf{B}$ applied in two (orthogonal) directions with respect to the $\textbf{c}$ axis of GaN : parallel $\textbf{B} || \textbf{c}$  ($M_{||}$) and perpendicular $\textbf{B} \perp \textbf{c}$ ($M_{\perp}$). The obtained results show, that at low to moderate fields, $|B| < B_{SORT1}$, a significant magnetic anisotropy with easy axis perpendicular to the $\textbf{c}$ axis is observed ($|M_{\perp}|$ > $|M_{||}|$). However for  $B_{SORT1} > |B| > B_{SORT2}$ the magnetic anisotropy seems to reverse its sign, that is $|M_{\perp}|$ < $|M_{||}|$. However, on increasing $B$ much further the second SRT-like effect appears, resulting in $|M_{\perp}|$ being again larger that $|M_{||}|$ for $|B| > B_{SORT2}$. It is found that the value of $B_{SORT1}$ depends on the crystal field model parameters and the number of ions $N$ in a given cluster,  whereas $B_{SORT2}$ is controlled mostly by the strength of the spin-orbit coupling $\lambda$. We show that the observed effect arises from the interplay of the crystalline environment and spin–orbit interaction. The explanation is given in terms of the spin $M_S$ and orbital momentum $M_L$ contributions to the total magnetization $M = M_S+M_L$. Detailed analysis show that the magnetic anisotropy of spin and orbital momentum components is unchanged , that is at any given $B$ we have $|M_{S,\perp}|$ > $|M_{S,||}|$ and $|M_{L,\perp}|$ > $|M_{L,||}|$. In particular, we show that in the low magnetic field region $|B| < B_{SORT1}$, MA is controlled by the dominant spin component $M_S$, whereas for higher magnetic fields $|B| > B_{SORT1}$ (in the spin saturation regime) MA depends on values of $M_L$. Therefore we refer to this effect as spin orbital reorientation transition (SORT).

It is worth noting, that (Ga,Mn)N has been intensively investigated both experimentally \cite{Gosk:2005_PRB, Sztenkiel:2016_NatComm, Gas:2018_JALCOM, Stefanowicz:2010_PRB, Bonanni:2011_PRB} and theoretically \cite{Gosk:2005_PRB, Sztenkiel:2016_NatComm, Sztenkiel:2020_NJP, Edathumkandy:2021_arXiv}, with CFM results reproducing quantitatively the experimental data. Recently, we have also shown that in (Ga,Mn)N-based heterostructures the magnetic anisotropy can be controlled by an electric field $E$ owing to the strong coupling between piezoelectricity and magnetism in this material \cite{Sztenkiel:2016_NatComm}. By piezoelectric modification of the $c$-axis lattice parameter we controlled the Mn$^{3+}$ single ion magnetic anisotropy in GaN. In the ferromagnetic state of (Ga,Mn)N, the electric field control of MA leads to decrease of coercive field $H_C$, and a non-reversible magnetization switching for magnetic fields close to $H_C$. It is of interest then to study magnetic properties of this material in more detail and investigate the effects related to the MA modification by other parameters that $E$.

\section{Crystal field model}

In this paper, the magnetization of small Mn$^{3+}$ magnetic clusters in the wurtzite GaN diluted magnetic semiconductor is probed using crystal field model \cite{Vallin:1974_PRB, Gosk:2005_PRB, Tracy:2005_PRB, Stefanowicz:2010_PRB}. Using the CFM approach it was possible to explain the magnetic \cite{Gosk:2005_PRB, Stefanowicz:2010_PRB, Bonanni:2011_PRB}, magnetoelectric \cite{Sztenkiel:2016_NatComm} and magnetooptic \cite{Wolos:2004_PRB_b}  experimental results in dilute (Ga,Mn)N.

The electronic structure of single Mn$^{3+}$ ion in GaN can be described by the following Hamiltonian \cite{Wolos:2004_PRB_b, Sztenkiel:2016_NatComm, Sztenkiel:2020_NJP, Edathumkandy:2021_arXiv}

\begin{equation}
\mathcal{H}(j)=\mathcal{H}_{CF}+\mathcal{H}_{TR}+\mathcal{H}_{JT}(j)+\mathcal{H}_{SO}+\mathcal{H}_{B}
\end{equation}

Here, the magnetocrystalline anisotropy is represented by the cubic crystal field $\mathcal{H}_{CF}=-\frac{2}{3}B_4(\hat{O}_4^0-20\sqrt{2}\hat{O}_4^3)$, the trigonal distortion along the hexagonal $\mathbf{c}$-axis of GaN  $\mathcal{H}_{TR}=B_2^0\hat{O}_4^0+B_4^0\hat{O}_4^2$, and the static Jahn-Teller distortion $\mathcal{H}_\textrm{JT}(j)=\tilde{B}_2^0\hat{\Theta}_4^0+\tilde{B}_4^0\hat{\Theta}_4^2$. These terms are expressed by the equivalent Stevens operators $\hat{\Theta}$ and  $\hat{O}$ for tetragonal distortion along one of the Jahn-Teller (J-T) cubic axis $e_\textrm{JT}^j$ ($j=A,B,C$) and the trigonal axis $\left[001\right] \parallel \mathbf{c}$ respectively. The J-T vectors are defined as follows: $\scriptstyle \textbf{e}^A_\textrm{JT}=\left[{\sqrt{2/3},0,\sqrt{1/3}}\right]$, $\scriptstyle \textbf{e}^B_\textrm{JT}=\left[{-\sqrt{1/6},-\sqrt{1/2},\sqrt{1/3}}\right]$, $\scriptstyle \textbf{e}^C_\textrm{JT}=\left[{-\sqrt{1/6},\sqrt{1/2},\sqrt{1/3}}\right]$. The CFM approach is supplemented with the spin orbit coupling $\mathcal{H}_{S0}=\lambda \hat{\textbf{L}}\cdot\hat{\textbf{S}}$ and the Zeeman term $\mathcal{H}_B=\mu_B(g_L\hat{\textbf{L}}+g_S\hat{\textbf{S}})\textbf{B}$, where g-factors are $g_S=2$, $g_L=1$ and $\mu_B$ represents the Bohr magneton. The numerical values of CFM parameters are summarized in Tab.~\ref{table:CFMParameters}. 

For Mn$^{3+}$ ion ($S=2$ and $L=2$), the basis states are represented by $|m_L,m_S>$ with respective spin and orbital angular momentum quantum numbers $-2 \leq m_S \leq 2$ and  $-2 \leq m_L \leq 2$. Then, the eigen-values $E_i$ and eigen-functions $|\phi_i>$, obtained by diagonalization of $25$ x $25$ Hamiltonian matrix, serve to calculate the average paramagnetic moment $\textbf{M}$ of Mn ion according to the following formulas

\begin{equation}
\textbf{M}=\frac{1}{Z}\sum_{j=A, B, C}(Z_j \textbf{m}^j)
\label{eq:MZ}
\end{equation}

\begin{equation}
\textbf{m}^j=\frac{-\mu_B\sum_{i=1}^{25}<\phi_i^j| g_L \hat{\textbf{L}}+g_S\hat{\textbf {S}}|\phi_i^j>exp(-E^j_i/k_{B}T)}{\sum_{i=1}^{25}exp(-E^j_i/k_{B}T)}
\label{eq:MSL}
\end{equation}

with $Z_j=\sum_{i}exp(-E^j_i/k_{B}T)$ being the partition function at the $j$-th JT center and $Z=\sum_{j=A, B, C}Z_j$. Due to the presence of three J-T directions, eigen-values $E^j_i$ and eigen-functions $|\phi_i^j>$ also depend on particular J-T configuration $j=A, B, C$. 

\begin{figure}[htp]
    \centering
    \includegraphics[width=4cm]{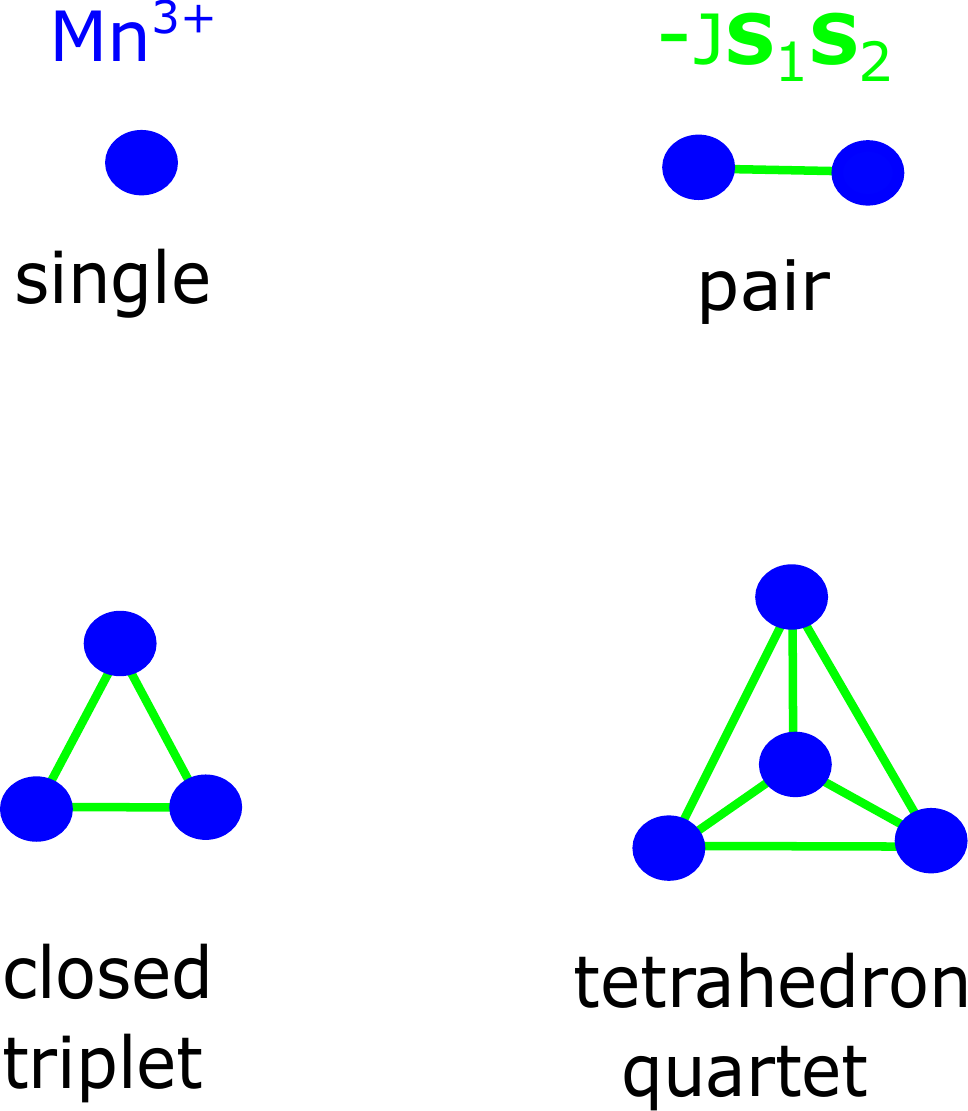}
    \caption{Selected clusters composed of magnetic ions coupled by a ferromagnetic superexchange interaction (green lines). Due to the computational complexity, the number of Mn$^{3+}$ ions in a cluster is limited to 4.}
    \label{fig:galaxy}
\end{figure}

Next, following Ref.~\onlinecite{Sztenkiel:2020_NJP, Edathumkandy:2021_arXiv}, we extended the single ion CFM approach, and include in the simulations clusters composed of Mn ions, coupled by a ferromagnetic superexchange interaction $\mathcal{H}_{Exch}=-\sum_{ [k,l] } J \hat{\textbf{S}}_{k} \cdot \hat{\textbf{S}}_{l}$. Here, $[k,l]$ represents all interacting Mn-Mn pairs within a given cluster. The value of superexchange coupling constant is set to $J = 10$~meV (c.f. Ref.~\onlinecite{Sztenkiel:2020_NJP, Edathumkandy:2021_arXiv} for details). As the size of Hamiltonian matrix grows very fast with the number of ions $N$ in a cluster ($25^N \times 25^N$), our simulations are limited to $N \leq 4$ \cite{Sztenkiel:2020_NJP}. The Mn$^{3+}$ cluster types investigated here are shown in Fig.~\ref{fig:galaxy}.

\begin{table}[h!]
\centering
\begin{tabular}{ p{1.1cm} p{0.9cm} p{1.1cm} p{1.1cm} p{1.1cm} p{0.9cm} p{0.9cm} p{0.1cm}} 
 \hline\hline
 \\
 \centering $B_{4}$  & \centering $B^0_{2}$  & \centering $B^0_{4}$  & \centering $\widetilde B^0_{2}$  & \centering $\widetilde B^0_{4}$  &  \centering  $\lambda_{TT}$  & \centering $\lambda_{TE}$ &  \\ [1ex] 
 
\hline
 \vspace{1mm}
 \\

 \centering 11.44  & \centering 2.2  & \centering -0.292 & \centering -5.85  & \centering -1.17  & \centering 5.5  & \centering 11.5 &   \\ 
 \hline\hline
\end{tabular} 
\caption{The crystal field model parameters of Mn$^{3+}$ ion in GaN. All values are in meV.}
\label{table:CFMParameters}
\end{table}

\begin{figure*}[htp]
    \centering
    \includegraphics[width=16cm]{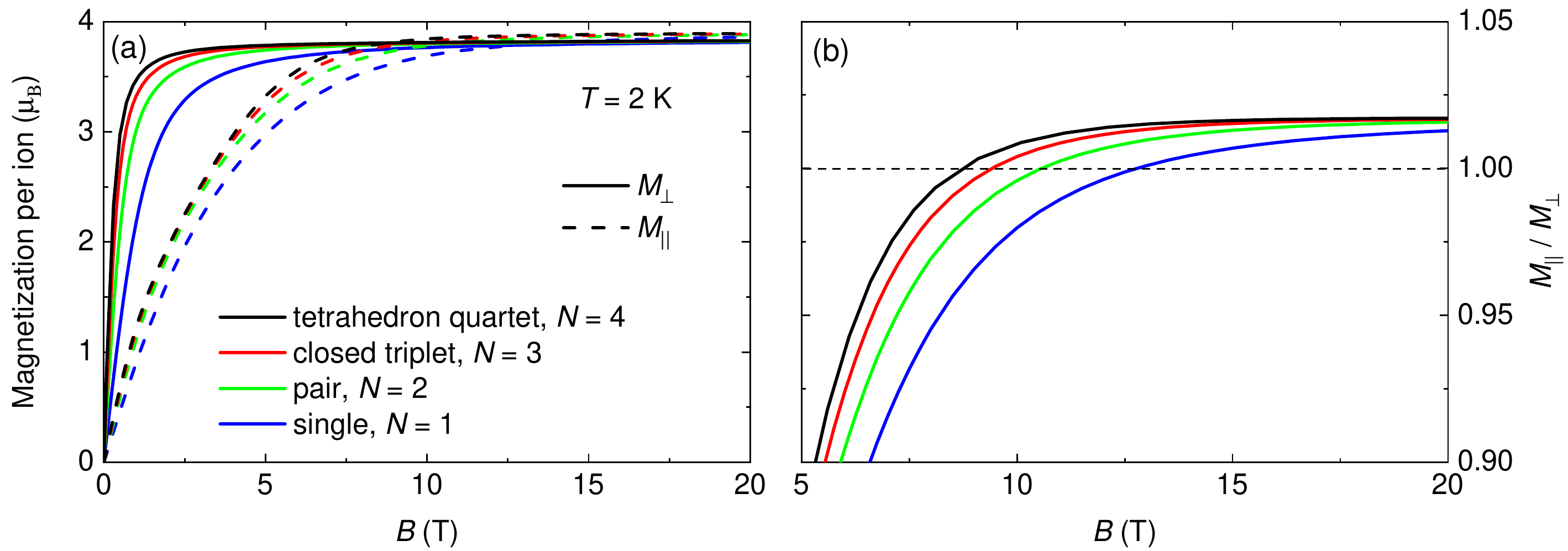}
    \caption{(\textbf{a}) Magnetization per one ion of selected Mn$^{3+}$ magnetic clusters in GaN. The Mn ions coupled by ferromagnetic superexchange interaction with $J=10$~meV, are still in the paramagnetic regime. The magnetization $M$ is computed as a function of the magnetic field $\textbf{B}$ applied in two orthogonal directions: parallel $\textbf{B} || \textbf{c}$ ($M_{||}$) and perpendicular $\textbf{B} \perp \textbf{c}$ ($M_{\perp}$) to the $\textbf{c}$ axis of GaN. (\textbf{b}) The $M_{||}/M_{\perp}$ ratio as a function of $B$. The SORT behavior takes place at magnetic fields $B_{SORT1}$ corresponding to $M_{||}/M_{\perp}=1$.}
    \label{fig:mag_SORT1}
\end{figure*}

\section{Numerical results}
\subsection{First spin orbital reorientation transition}

In our recent paper (Ref.~\onlinecite{Sztenkiel:2020_NJP}), we also studied the magnetization of small Mn$^{3+}$ clusters in GaN with the ferromagnetic superexchange interaction between constituting ions. We observed the presence of the effect similar to the spin reorientation transition at the magnetic field of $B_{SORT1} = ~6\div 9$~T ( labeled by $B_a$ in Ref.~\onlinecite{Sztenkiel:2020_NJP} ). Here we investigate this phenomenon in more detail, and provide its underlying explanation. We refer to this effect as to the spin orbital reorientation transition SORT (the reason for this will be explained later in the text). It should be pointed out that, to our best knowledge, the effect discussed here has not been observed experimentally in (Ga,Mn)N or in any other material yet, and therefore awaits its experimental verification.
In Fig.~\ref{fig:mag_SORT1} \textbf{a} we present the magnetization $M$ per one ion, computed as a function of the external magnetic field $\textbf{B}$ applied in two directions with respect to the $\textbf{c}$ axis of GaN, namely $\textbf{B} || \textbf{c}$ ($M_{||}$) and $\textbf{B} \perp \textbf{c}$ ($M_{\perp}$).
At weak to moderate fields, $0 < B < B_{SORT1}$, a significant magnetic anisotropy with the easy axis perpendicular to the $\textbf{c}$ axis is observed ($M_{\perp}$ > $M_{||}$). The magnetic anisotropy is predominately uniaxial, originating from the trigonal deformation along $\textbf{c}$. However, for $B > B_{SORT1}$, the magnetic anisotropy seems to reverse its sign, with $M_{\perp}$ < $M_{||}$. The actual values of $B_{SORT1}$, exemplified in Table~\ref{table:BSORT1}, depend on the number of ions $N$ in a given cluster and the crystal field model parameters (c.f. the values of  $B_{SORT1}$ labeled by $B_a$ in Ref.~\onlinecite{Sztenkiel:2020_NJP}). However the effect itself seems to be robust, as in Ref.~\onlinecite{Sztenkiel:2020_NJP} we used different set of CFM parameters, what resulted in similar qualitative behavior.

\begin{table}[htb]
\centering
\begin{tabular}{ p{1.8cm} | p{1.0cm} p{1.0cm} p{1.0cm} p{1.0cm} p{0.1cm}} 
 \hline\hline
	\\

 \centering $N$  &  \centering  1   &  \centering 2 &  \centering 3 &  \centering 4 & \\
 
 \hline
 \vspace{1mm}
 \\
  \centering $B_{SORT1}$~(T)   &  \centering 12.8   &  \centering 10.5   &  \centering 9.4  & \centering  8.7 &      \\  
 \hline\hline
\end{tabular} 
\caption{Magnitudes of $B_{SORT1}$ as a function of number of ions $N$ in the cluster at temperature $T=2$~K.}
\label{table:BSORT1}
\end{table}

It is worth noting that the SORT effect is small, but clearly visible in numerical data presented in Fig.~\ref{fig:mag_SORT1}. To put our findings into perspective, assuming a volume of the magnetic layer of $V = 4\cdot10^{-6}$~cm$^{3}$ (5mm $\times$ 4mm $\times$ 200nm) and Mn concentration $x=0.01$ in Ga$_{1-x}$Mn$_x$N, we can estimate the difference of magnetizations at high magnetic field $B=15$~T ($B>B_{SORT1}$) and $T=2$~K as $M_{||} - M_{\perp} \cong 4\cdot10^{-7}$~emu. The estimated value is well above the SQUID detection limit of  $10^{-8}$~emu \cite{Gas:2019_MST}. However, in the usual experimental conditions the investigated magnetic thin film is grown on a bulky substrate. Then, using for example, standard superconducting quantum interference device (SQUID) magnetometer, a dedicated in situ compensation of the (sapphire) substrate has to be used \cite{Gas:2019_MST} to resolve this signal from the otherwise vastly dominating signal of the bulky substrate \cite{Gas:2021_JALCOM}.
Detailed survey on other sources of possible artifacts during magnetic measurements were reported in Ref.~\onlinecite{Sawicki:2011_SemSciTech, Gas:2019_MST, Gas:2021_JALCOM}. All this suggest, that the experimental observation of SORT effect in (Ga,Mn)N or in any alternative  material may cause some difficulties. 

\begin{figure}[htp]
    \centering
    \includegraphics[width=7.5cm]{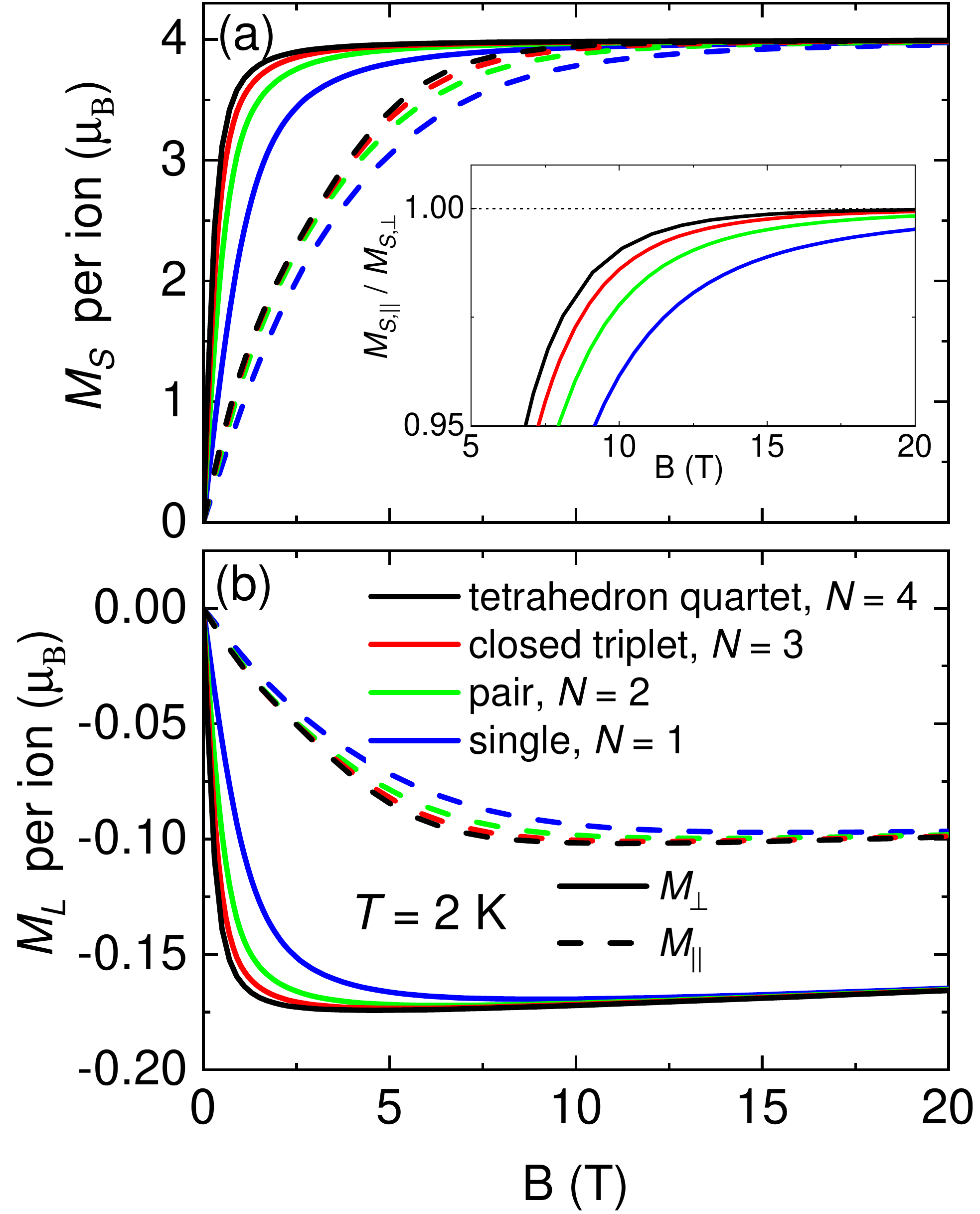}
    \caption{The spin $M_S$ (\textbf{a}) and orbital momentum $M_L$ (\textbf{b}) contributions (per one ion) to the total magnetization $M = M_S + M_L$. These contributions are computed as the thermodynamical and configurational average of the respective magnetic moment operators $ -\mu_B g_S\hat{\textbf {S}}$ and $-\mu_B g_L \hat{\textbf{L}}$ (see Eq.~\ref{eq:MZ} and \ref{eq:MSL}). $M_S$ and $M_L$ are computed as a function of the magnetic field $\textbf{B}$ applied in two directions with respect to the $\textbf{c}$ axis of GaN, namely $\textbf{B} || \textbf{c}$ ($M_{S, ||}$ and $M_{L, ||}$) and $\textbf{B} \perp \textbf{c}$ ($M_{S, \perp}$ and $M_{L, \perp}$). Antiparallel alignment of $M_S$ and $M_L$ is due to the presence of spin-orbit interaction $\lambda \hat{\textbf{L}}\cdot\hat{\textbf{S}}$ with $\lambda > 0$.}
    \label{fig:magSL_SORT1}
\end{figure}

\begin{figure}[htp]
    \centering
    \includegraphics[width=7.0cm]{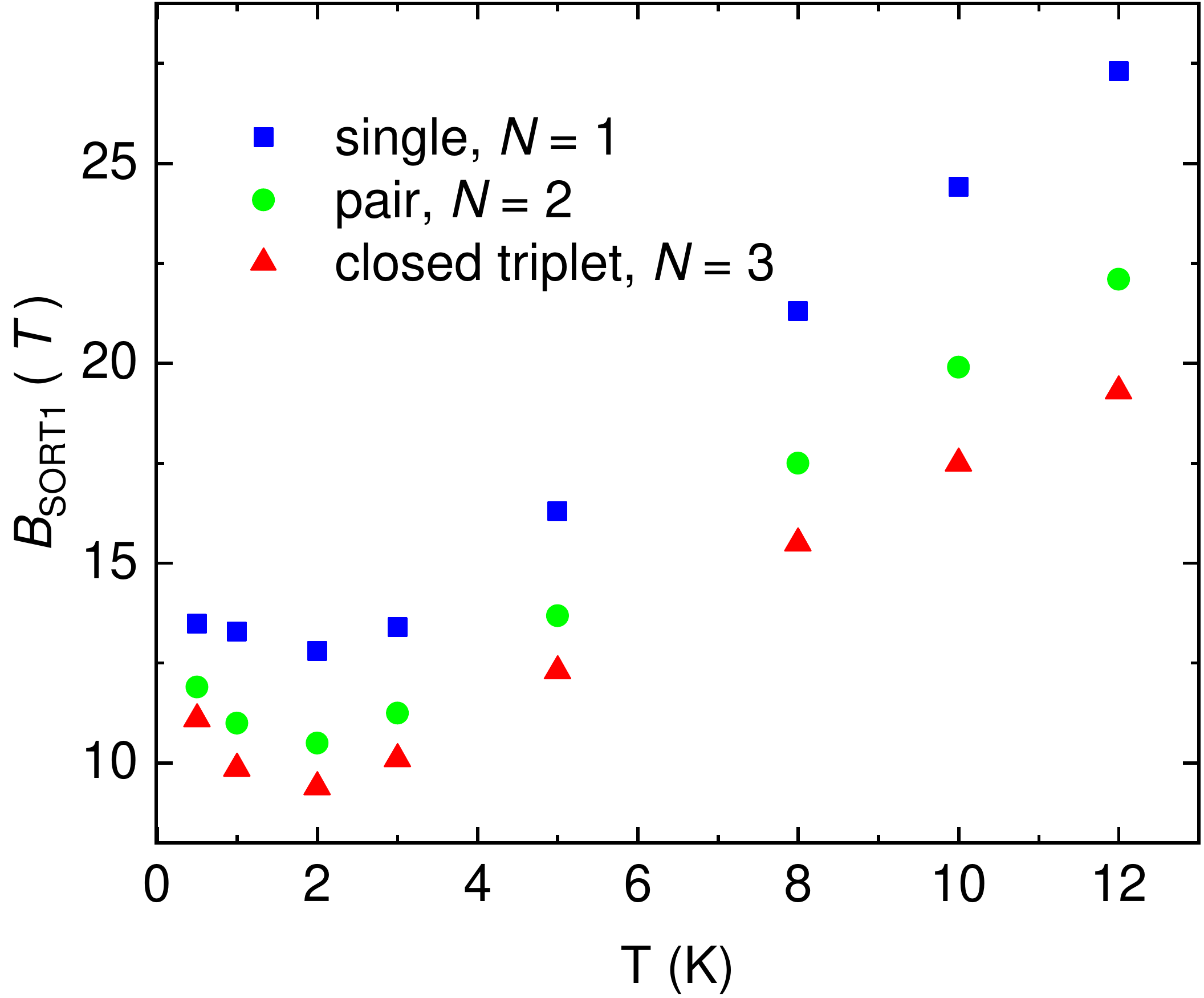}
    \caption{Values of $B_{SORT1}$ as a function of temperature. Only clusters with up to 3 ions are shown.}
    \label{fig:BSORT1vsT}
\end{figure}

\begin{figure}[htp]
    \centering
    \includegraphics[width=8.5cm]{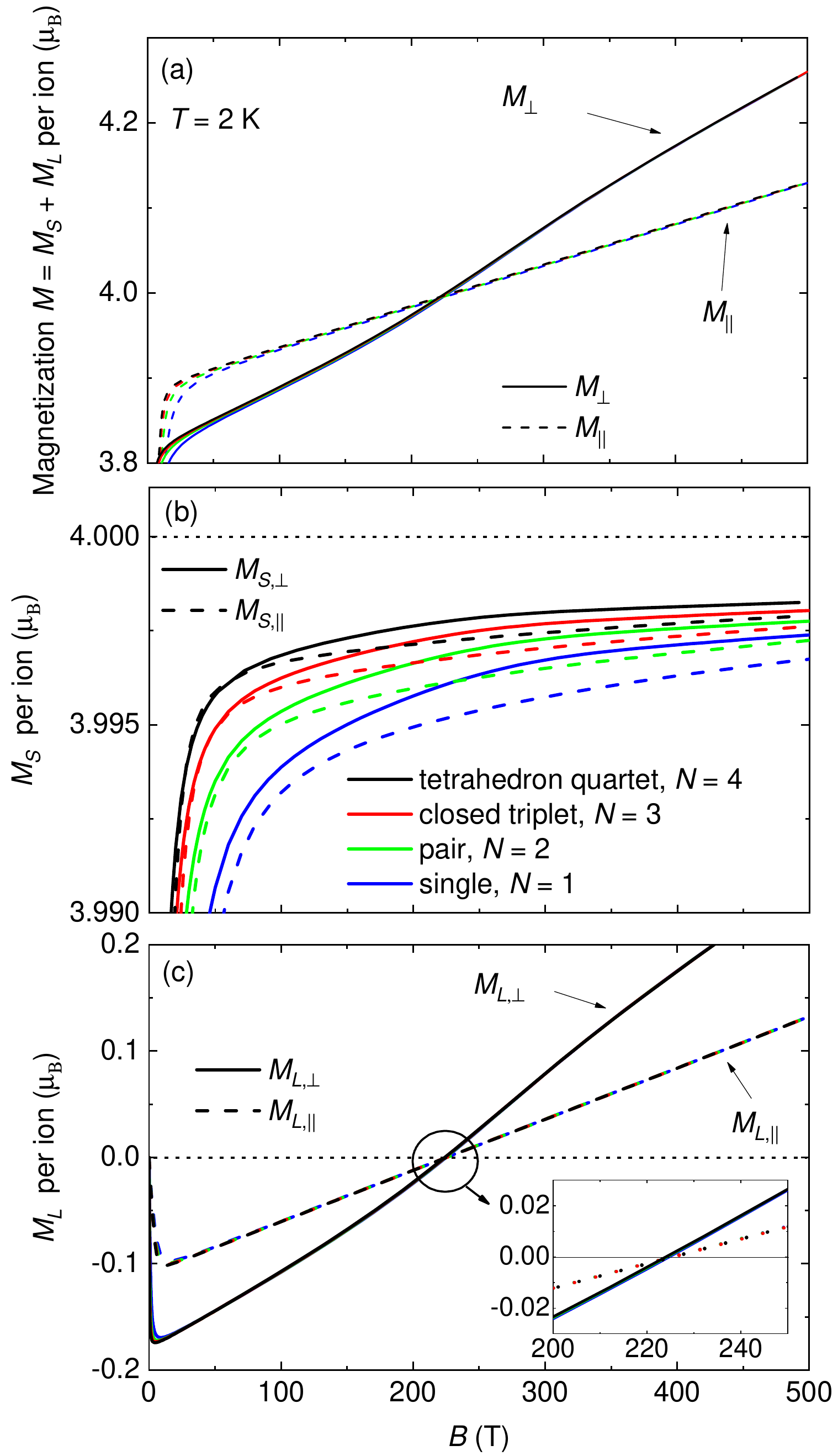}
    \caption{(a) The total magnetization per one ion $M = M_S + M_L$, (b) the spin $M_S$ and (c) the orbital momentum $M_L$ contributions (per one ion) as a function of $B$. The magnetic field $\textbf{B}$ is applied in two directions with respect to the $\textbf{c}$ axis of GaN, namely $\textbf{B} || \textbf{c}$ ($M_{||}$, $M_{S, ||}$ and $M_{L, ||}$) and $\textbf{B} \perp \textbf{c}$ ($M_{\perp}$, $M_{S, \perp}$ and $M_{L, \perp}$).}
    \label{fig:magSL_SORT2}
\end{figure}

The simple explanation of SRT is that, the magnetic easy axis is in the c-plane at medium magnetic fields $B$ and it reorients into the $c$-axis upon increasing $B$ across $B_{SORT1}$. However we claim, that in the presented case, the magnetic anisotropy is unchanged for high magnetic fields ($|B| > B_{SORT1}$). The magnetic easy axis is always perpendicular to the $c$ axis of GaN, regardless of the magnitude of $B$. Detailed analysis show that the observed SORT effect arises from the interplay of the crystalline environment and spin–orbit coupling. As shown in eq.~\ref{eq:MZ} and \ref{eq:MSL}, $M$ is computed as the thermodynamical and configurational average of total magnetic moment operator $-\mu_B( g_L \hat{\textbf{L}}+g_S\hat{\textbf {S}})$. Therefore, the explanation of SORT behavior, is given in terms of the spin $M_S$ and orbital momentum $M_L$ contributions to the total magnetization $M = M_S + M_L$. In Fig.~\ref{fig:magSL_SORT1} we plotted the $M_S$ and $M_L$ as a function of magnetic field. 

 The crystal field interaction acts on the orbital angular momentum $\textbf{L}$ degrees of freedom resulting in the quenching of $\textbf{L}$ and development of the single-ion magnetic anisotropy. The resulting magnetic anisotropy in $M_L$ is transferred to the spin part $M_S$ by the spin-orbit interaction. As a result, the values of $M_L$ are rather low as compared to $M_S$ (due to the quenching of $L$), but the magnetic anisotropy of $M_L$ is much more pronounced, than that of $M_S$. Due to the positive value of spin-orbit parameter $\lambda > 0 $, $M_S$ and $M_L$ are anti-parallel. Additionally the Zeeman energy for the  spin part is much larger than for the orbital momentum part, what causes $M_S$ to be parallel to the direction of the magnetic field, with $M_L$ being negative for $B > 0$, as shown in fig.~\ref{fig:magSL_SORT1} (later we will show, that very large magnetic field $B > 220$~T can reorient $M_L$ and change its sign). It is worth noting that the magnetic easy axis is always perpendicular to the $c$ axis of GaN both for $M_S$ and $M_L$, that is $|M_{S, \perp}| > |M_{S, ||}|$ and $|M_{L, \perp}| > |M_{L, ||}|$. The important distinction is, that for $0 \leqslant B \leqslant 20$~T we have $M_{L, ||} > M_{L, \perp}$ and $M_{S, ||} < M_{S, \perp}$.

\begin{figure*}[htp]
    \centering
    \includegraphics[width=12cm]{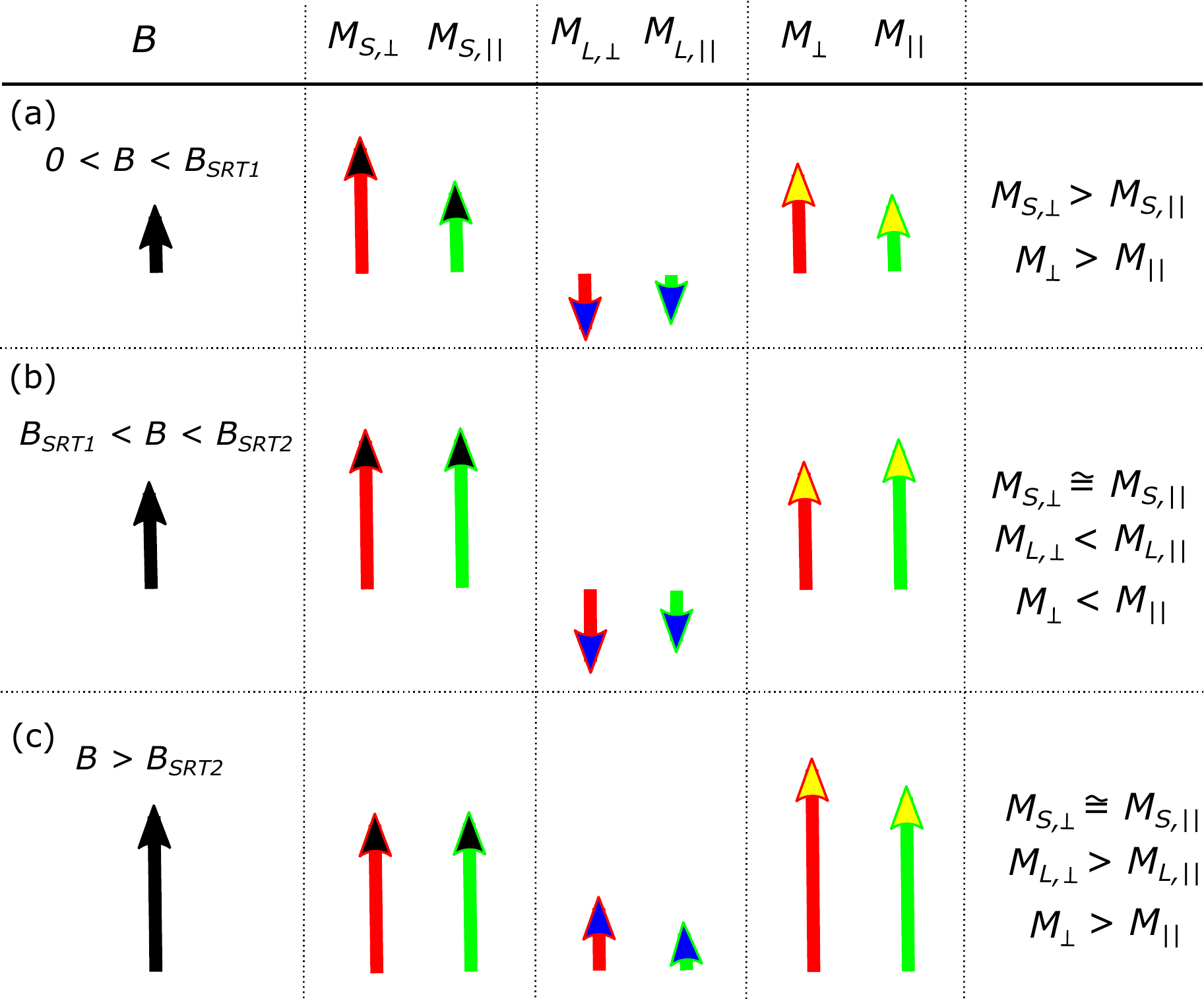}
    \caption{Schematic explanation of the observed spin reorientation like transitions. Black arrows correspond to the strength of the applied magnetic field $B$, where in (a) $0 < B < B_{SORT1}$, (b) $B_{SORT1} < B < B_{SORT2}$ and (c) $B > B_{SORT2}$. Red and green arrows denote the magnetizations $M_S$, $M_L$ and $M = M_S+M_L$ calculated for the magnetic field applied perpendicular and parallel to the $\textbf{c}$ axis of GaN respectively. (a) The magnetic anisotropy (MA) is controlled by the dominant spin component $M_S$. Antiparallel alignment of $M_S$ and $M_L$ is due to the presence of spin-orbit interaction $\lambda \hat{\textbf{L}}\cdot\hat{\textbf{S}}$ with $\lambda > 0$. (b) and (c) In the spin saturation regime with $M_{S,\perp} \cong M_{S,||} \cong 4$~$\mu_B$ per ion, MA depends on values of the orbital contribution $M_L$ to the total magnetization $M$. (c) A very strong magnetic field $B$ overcomes the influence of the spin-orbit interaction what results in the reversal of $M_L$ and the occurrence of the second spin reorientation like transition.} 
    \label{fig:arrows}
\end{figure*}

In fig.~\ref{fig:magSL_SORT1} a we see that $M_S$ per one ion saturates at high magnetic field without the presence of spin reorientation transition. The saturation value is very close to 4~$\mu_B$, as expected. In the inset to fig.~\ref{fig:magSL_SORT1}.\textbf{a} the ratio of spin contributions to the total magnetization for magnetic field applied parallel $M_{S, ||}$ and perpendicular $M_{S, \perp}$ to the $c$-axis of GaN is presented. $M_{S, ||} / M_{S, \perp} < 1$ in the whole magnetic field range $0 \leqslant B \leqslant 20$~T, indicating the absence of SRT. However, as shown in fig.~\ref{fig:mag_SORT1}, at high magnetic fields $M_{||} > M_{\perp}$, due to the fact that $M_{S, ||} \cong M_{S, \perp}$ and $M_{L, ||} > M_{L, \perp}$ (see fig.~\ref{fig:magSL_SORT1} b). In summary, up to $B_{SORT1}$ the magnetic anisotropy is controlled by the dominant spin component $M_S$ and then $M_{||} < M_{\perp}$. Contrary, above  $B_{SORT1}$, where $M_{S, ||} \cong M_{S, \perp}$, MA is governed by the orbital momentum contribution $M_L < 0$, with $M_{||} > M_{\perp}$. We underline here that, the presence of such SORT effect should also be observed in other materials with positive value of spin-orbit parameter $\lambda > 0 $, where both $M_S$ and $M_L$ are anti-parallel.

The SORT effect takes place at magnetic fields where $M_{S, \perp} \cong M_{S, ||}$. Now it is relatively easy to understand why $B_{SORT1}$ depends on the number of ions $N$ in a given cluster.  Due to the presence of the ferromagnetic coupling between atoms the magnitude of $M_S$ at given $B$ increases with $N$. Then, for larger clusters both $M_{S, \perp}$ and  $M_{S, ||}$ saturates at lower values of $B$, as shown in Fig.~\ref{fig:magSL_SORT1} \textbf{a}. Similarly, by decreasing the trigonal anisotropy constants ($B^0_{2}$  and  $B^0_{4}$), we can reduce the strenght of the uniaxial magnetic anisotropy. This will shift the saturation value of the hard spin magnetization $M_{S, ||}$ to lower values of $B$, resulting in smaller $B_{SORT1}$  (c.f. $B_{SORT1}$ values in Ref.~\onlinecite{Sztenkiel:2020_NJP} where different set of trigonal CFM parameters were used).

In Fig.~\ref{fig:BSORT1vsT} we plot magnitudes of $B_{SORT1}$ as a function of temperature. Interestingly, a non-monotonous behaviour is observed in $B_{SORT1}(T)$ curves. The minimum values of $B_{SORT1}$ are attained at $T\cong 2$~K. Then, for $T\ > 2$~K we observe a strong increase in $B_{SORT1}$ with temperature.

\subsection{Second spin orbital reorientation transition}

In fig.~\ref{fig:magSL_SORT2} we present $M$, $M_S$ and $M_L$ in a very high magnetic field range $0 \leqslant B \leqslant 500$~T (we are aware, that such high values of $B$ are not available in the standard magnetic experimental set-ups, in which usually $|B| \leqslant 7$~T). The obtained data reveals the presence of the second SORT effect at $B_{SORT2} \cong 222$~T. In fig.~\ref{fig:magSL_SORT2} \textbf{b}, we see that $M_S$ is large, very close to the saturation value of 4~$\mu_B$. However, $M_L$ varies strongly with the magnetic field. By increasing $B$ we increase the Zeeman interaction related to the orbital momentum $\mathcal{H}_B^L=\mu_B(g_L\hat{\textbf{L}})\textbf{B}$, what drives $M_L$ toward positive value. At the same time, the spin orbit-coupling $\mathcal{H}_{S0}=\lambda \hat{\textbf{L}}\cdot\hat{\textbf{S}}$ maintains a negative value of $M_L$. At magnetic field of $B_{SORT2}$, the Zeeman related $\textbf{B}$ and spin-orbit related $\frac{\lambda \hat{\textbf{S}}}{\mu_B g_L}=-\frac{\lambda \hat{\textbf{M}_S}}{\mu_B^2 g_L g_S}$ magnetic field acting on $\hat{\textbf{L}}$ cancel, what results in $M_L = 0$. Then, for $M_S$=4$\mu_B$ (at large $B$ we are in the spin saturation regime) and $\lambda = 5.5$~meV we have $B_{SORT2}=\frac{\lambda \hat{\textbf{M}_S}}{\mu_B^2 g_L g_S} \cong 224$~T. This analytical value of $B_{SORT2}$ agrees very well with the numerical one, obtained from data presented in fig.~\ref{fig:magSL_SORT2} \textbf{c}. It seems that $B_{SORT2}$ depends only on $\lambda$ and the value of ${\textbf{M}_S}$ per one ion at $B=B_{SORT2}$, and is very weakly dependent on the number of ions $N$ or other CFM parameters. It is worth noting that, in the case of small spin-orbit interaction, one can infer the value of $\lambda$ from experimental $M(B)$ curves measured in two perpendiculars direction using the following relation $\lambda = \mu_B g_L B_{SORT2} / S$. Finally, in fig.~\ref{fig:arrows} we present schematic explanation of the observed SORT effects in (Ga,Mn)N.

In the calculations presented above, we used $\lambda = \lambda_{TT}$. The reason behind this is as follows. The $^5$D state of the free Mn$^{3+}$ ion is splitted by the cubic crystal field $H_{\mathrm{CF}}$ into a tenfold orbital doublet $^5$E and 15-fold orbital triplet $^5$T. In general, the hybridization of the $^5$E state with the ligand wave functions is different than for $^5$T, resulting in three distinct spin-orbit coupling constants $\lambda_{TT}$, $\lambda_{TE}$ and $\lambda_{EE}=0$ (corresponding to various combination of $^5$E and $^5$T). As $^5$T is the ground state, at low temperatures only this state is practically occupied.

As stated before, for practical reasons, the simulations presented here are restricted to clusters composed of up to four interacting ions. In general, it is possible to bypass this limitation and model very large systems using atomistic spin model supplemented with the stochastic Landau-Lifshitz-Gilbert equation (sLLG) \cite{Evans:2015_PRB, Evans:2014_JPhysCM, Edathumkandy:2021_arXiv}. Recently, we have performed comparative study of magnetization of small Mn$^{3+}$ clusters in GaN using classical (sLLG) and quantum mechanical (CFM) approaches \cite{Edathumkandy:2021_arXiv}. We have found that classical simulations reproduce very well the quantum magnetization curves at low temperatures or in the strong coupling regime. However, the sLLG approach is based only on an effective spin Hamiltonian. The inclusion of orbital angular momentum degrees of freedom and spin-orbit interaction in the classical simulations is clearly difficult. A particular challenge is related to proper reproduction of the effect of quenching of the orbital angular momentum. So the conclusion is that exact CFM quantum mechanical approach is indispensable in special cases such as those presented in this paper. 

\subsection{Outlook}

In general, spin orbital reorientation transition should be observed in other materials which are characterized by positive value of spin-orbit parameter $\lambda > 0 $, a non zero value of the orbital momentum $L$ and at least single ion uniaxial magnetic anisotropy. In order to observe this effect within the standard range of experimentally available external magnetic fields $|B| < 7$~T, two conditions have to be met. Firstly, the investigated material should have a weak single ion magnetocrystalline anisotropy (small value of MA will shift the saturation value of the hard spin magnetization $M_S$ to lower values of $B$ ). Secondly, as $B_{SORT2}$ depends linearly on the magnitude of $\lambda$, we search for materials with weak spin-orbit coupling. In our opinion there are plenty systems, both bulky or 2D materials, in which SORT can be experimentally detected. Especially, the amount of research of 2D magnets has surged significantly  in recent years. In Ref.~\onlinecite{Jiang:2021_ApplPhysRev} a list of more that one hundred 2D magnets were presented with their key magnetic properties such as magnetic moment per transition metal ion, Curie temperature $T_\mathrm{C}$ or magnetocrystalline anisotropy energy (MAE) per unit cell. Many of these materials, characterized by small values of MAE, can form a suitable platform for experimental observation of the SORT effects.

\section{CONCLUSIONS}

In this paper, we numerically computed the magnetization $M(B,T=2$~K$)$ of small Mn$^{3+}$ clusters in GaN being in the paramagnetic regime. The magnetic field was applied in two perpendicular directions, namely $\textbf{B} || \textbf{c}$ ($M_{||}$) and $\textbf{B} \perp \textbf{c}$ ($M_{\perp}$). In particular, we observe two spin reorientation like transitions in $M(B)$ curves, at magnetic fields $B_{SORT1}$ and $B_{SORT2}$. That is, at low to moderate fields, $|B| < B_{SORT1}$, a significant magnetic anisotropy with easy axis perpendicular to the $\textbf{c}$ axis of GaN  is observed ($|M_{\perp}|$ > $|M_{||}|$). At an intermediate fields, $ B_{SORT1} > |B| > B_{SORT2}$, the magnetic anisotropy seems to reverse its sign, where $|M_{\perp}|$ < $|M_{||}|$. However, increasing $|B|$ further, leads to the second SRT-like effect, resulting in $|M_{\perp}|$ being again larger that $|M_{||}|$ for $|B| > B_{SORT2}$. We show that a simple explanation of the observed effect as being a consequence of reorientation of the magnetic easy axis upon increasing $B$ is invalid. The explanation of this effect is given in terms of the spin $M_S$ and orbital momentum $M_L$ contributions to the total magnetization $M = M_S + M_L$. Therefore, we refer to this effect as spin orbital reorientation transition (SORT). We demonstrate that the magnetic anisotropy of spin and orbital momentum components is unchanged for high magnetic fields, and the easy axis of $M_S$ and  $M_L$ is always perpendicular to the $\textbf{c}$-axis of GaN. In particular at given $B$ we have $|M_{S,\perp}|$ > $|M_{S,||}|$ and $|M_{L,\perp}|$ > $|M_{L,||}|$ in the whole studied magnetic field range. Additionally, we show that in the low magnetic field region $|B| < B_{SORT1}$, magnetic anisotropy is controlled by the dominant spin component $M_S$, whereas for higher magnetic fields $|B| > B_{SORT1}$ (in the spin saturation regime) MA depends on values of $M_L$. Detailed analysis shows, that $B_{SORT1}$ depends on material parameters, temperature and the number of ions $N$ in the cluster. On the other hand, $B_{SORT2}$ is controlled solely by the strength of the spin-orbit coupling $\lambda$ and the value of ${\textbf{M}_S}$ per one ion at $B=B_{SORT2}$. Finally, we claim that the SORT effect should be present in other materials with not completely quenched (non zero) orbital angular momentum $L$, possessing a uniaxial magnetic anisotropy and the positive value of spin-orbit coupling $\lambda > 0$. 






\section*{Acknowledgments}

We would like to thank M. Sawicki for discussion and valuable suggestions. The work is supported by the National Science Centre (Poland) through project OPUS 2018/31/B/ST3/03438. The calculations were made with the support of the Interdisciplinary Center for Mathematical and Computational Modeling of the University of Warsaw (ICM UW) under the computational grant no GB77-6.





\end{document}